\documentclass[aip,reprint,amsfonts,amssymb,amsmath,showpacs,longbibliography]{revtex4-1}

\usepackage{graphicx}
\usepackage{dcolumn}
\usepackage{bm}
\usepackage{amssymb}
\usepackage{epstopdf}
\usepackage{color}

\begin{document}

\title{Compact cold atom clock for on-board timebase: tests in reduced gravity}

\author{Mehdi Langlois}
\email{m.langlois@bham.ac.uk}
\affiliation{SYRTE, Observatoire de Paris, PSL Research University, CNRS, Sorbonne Universit\'{e}s, UPMC Univ. Paris 06, LNE, 61 avenue de l'Observatoire, 75014 Paris, France}

\author{Jean-Fran\c{c}ois Schaff}
\affiliation{Muquans, Institut d'Optique d'Aquitaine, rue Fran\c{c}ois Mitterand, 33400 Talence, France}

\author{Luigi De Sarlo}
\affiliation{SYRTE, Observatoire de Paris, PSL Research University, CNRS, Sorbonne Universit\'{e}s, UPMC Univ. Paris 06, LNE, 61 avenue de l'Observatoire, 75014 Paris, France}

\author{Simon Bernon}
\affiliation{LP2N, Institut d'Optique d'Aquitaine, Universit\'{e} Bordeaux, CNRS, 33400 Talence, France}

\author{David Holleville}
\email{david.holleville@obspm.fr}
\affiliation{SYRTE, Observatoire de Paris, PSL Research University, CNRS, Sorbonne Universit\'{e}s, UPMC Univ. Paris 06, LNE, 61 avenue de l'Observatoire, 75014 Paris, France}

\author{No\"{e}l Dimarcq}
\affiliation{SYRTE, Observatoire de Paris, PSL Research University, CNRS, Sorbonne Universit\'{e}s, UPMC Univ. Paris 06, LNE, 61 avenue de l'Observatoire, 75014 Paris, France}

\date{\today}

\begin{abstract}
We present a compact atomic clock using cold rubidium atoms based on an isotropic light cooling, a Ramsey microwave interrogation and an absorption detection. Its technology readiness level is suitable to industrial transfer. We use a fibre optical bench, based on a frequency-doubled telecom laser. The isotropic light cooling technique allows us to cool down the atoms in 100~ms and works with a cycle time around $200$~ms. We carried out measurements in simulated microgravity and obtained the narrowest fringes ever recorded in microgravity.
\end{abstract}

\maketitle

\section{Introduction}
%
%\subsection{Beyond GNSS and H maser}

It is well known that global navigation satellite systems (GNSS) such as GPS can provide synchronisation to UTC better than 40~ns. This limit however is typically reached only for a stationary platform with a calibrated receiver, for a moving platform the timebase provided by the GNSS is subject to more systematics including service availability and reliability. Furthermore there is an increasing number of platforms for which high accuracy intertial navigation is required and GNSS is not an option. Examples of these platforms are submarines and deep-space missions. Last but not least, a highly reliable and accurate timebase could be used to upgrade the existing facilities on board the satellites of GNSS constellations.

The key ingredient of autonomous timebase generation is an oscillator which can provide an intrinsically high stability ($1~\mu$s over one year or $3 \times 10^{-14}$ of relative instability \cite{1995Bhaskaran}). This kind of performance is currently available only using hydrogen masers which have indeed been miniaturized and constitute the main on board the satellites of GALILEO european GNSS.

%\subsection{Cold atoms clocks}

Cold atom based atomic clocks currently realize the most accurate primary frequency standard in several metrology institutes worldwide \cite{6174184} and will also be on board the international space station thanks to the PHARAO clock \cite{Laurent2006}. 

Despite those great achievements, no onboardable cold-atom based clock capable of achieving similar performances to hydrogen masers in an easier configuration than the PHARAO clock has ever been demonstrated.

%\section{Rubiclock}

In this paper, we describe a compact atomic clock based on cold $^{87}$Rb atoms which will fill this gap. To this end we report on the clock operation onboard an airbus A300 performing parabolic flights and we analyze the limits of the clock in the present and ideal configuration for operation on board both in standard and reduced gravity. We compare the results obtained to those on ground.

%\subsection{Outline?}

\section{Setup}

\subsection{Approach}

Our setup is a full redesign of our previous experiment \cite{PhysRevA.82.033436}, with the following major differences: we now use $^{87}$Rb atoms instead of $^{133}$Cs and the laser system is based on the frequency-doubled telecom technology \cite{Lienhart2007}, allowing very compact and robust systems. The electronics, vacuum chamber and magnetic shielding have also been renewed and adapted to operation on board.

Using $^{87}$Rb is favorable for two reasons: the cold-collision shift is reduced compared to $^{133}$Cs \cite{RevModPhys.71.1}, allowing improved long-term stability, and it allows using compact fibre lasers.

\subsection{Laser system}

Our new laser system, produced by the company Muquans, is based on two fibre telecom external cavity diode lasers operating at 1560~nm, which are then frequency-doubled to reach the Rubidium lines around 780~nm. The first (master) is frequency-doubled and locked by absorption spectroscopy to the $^{85}$Rb $|5S_{1/2},F=3\rangle \rightarrow |5P_{3/2},F'=3,4\rangle$ crossover. A slave laser is phase locked to the master with an offset frequency $\sim 620$\,MHz provided by a direct digital synthesizer (DDS), allowing to change the frequency dynamically during the different stages. The repumper frequency is generated by a phase modulator operating at $\sim 6.6$~GHz. Power is provided by an Erbium-doped fibre amplifier. An acousto-optic modulator is used to reduce or turn off the power. A second frequency-doubling crystal is used after amplification, and a free-space splitter divides the power into two fibres (one for the cavity beams, and one for cooling and detection along the vertical direction). The power ratio was optimized to $\sim 30$~\% in the vertical beam with a waveplate to maximize the atom number. Inside the splitter, mechanical shutters are placed on the two paths to provide attenuation higher than $120$~dB during the Ramsey spectroscopy. The total output power was always $> 120$~mW at 780~nm. This full laser system (optics and electronics) is fully integrated into a 19-inch wide, 4~U high rack.

%The repumper is generated using an electro-optic modulator (EOM) which creates sidebands on the cooling beam. The modulation is controlled by subtracting the signal of a DDS to a 7 GHz signal from a microwave frequency chain. The preparation is done by depumping with the $|5S_{1/2},F=2\rangle \rightarrow |5P_{3/2},F’=2\rangle$ transition. We separate the Zeeman sublevels with a 10 mG magnetic field generated by a solenoid, but we do not make a Zeeman selection, so we keep about 30 \% of the atoms in the  $|5S_{1/2},F=1,m_{F}=0\rangle$ level. The detection is made by absorption with a retroreflected vertical probe beam through the atomic cloud, at $I_{sat}/5$ (for $^{87}$Rb D$_{2}$ transition $I_{sat}=1.669$ mW.cm$^{-2})$. The beam passes through a 7 mm aperture. The master/slave beat signal that is used to control the cooling, preparation and detection frequencies is generated by another channel of a DDS.

%of the HORACE clock in which the preparation of the atomic sample and the interrogation take place in the same physical region thereby drastically simplyfing the physics package. Furthermore the use of Rb atoms allows us to use an optical bench for atom manipulation which is based on telecom qualified components. The optical bench used in the experiments reported in this paper has been designed and realized by the French company Muquans.

\section{Experimental cycle}

\begin{figure}[h!]
	\includegraphics[width=\linewidth]{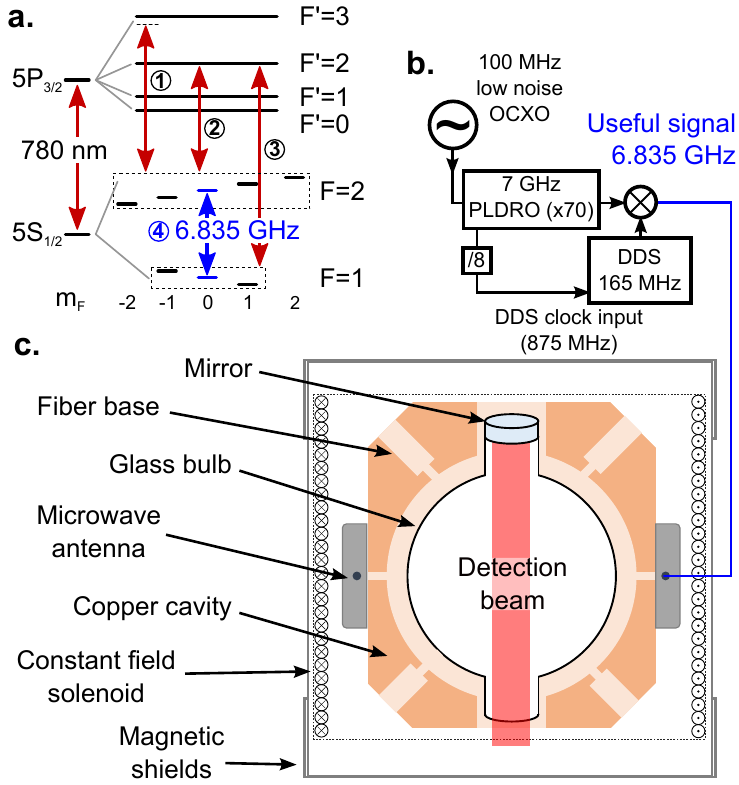}
	\caption{Experimental setup. \textbf{a.} Optical transitions involved in the detection and preparation of the cold sample with isotropic cooling (red): (1) cooling is performed red-detuned from the cycling transition, detection is done on resonance, (2) depumping to $F = 1$ is achieved on the $2 \rightarrow 2$ transition, (3) repumping light is obtained by phase modulation. (4, blue) Microwave clock transition. \textbf{b.} The clock frequency $\nu_c \simeq 6.835$~GHz is generated from a 100~MHz oven controlled crystal oscillator (OCXO), which is multiplied by 70 and then mixed to a signal generated by a DDS. This signal is injected through one of the two microwave antenna. \textbf{c.} The glass bulb is within a copper cavity which is enclosed in a solenoid generating a constant magnetic field. All these components are isolated from external magnetic fields by magnetic shields.}
	\label{fig_setup}
\end{figure}

As in our previous work, cooling, state preparation, interrogation and detection are carried out in a glass cell positioned at the centre of a spherical copper cavity. This is illustrated in Fig.~\ref{fig_setup}.

\subsection{Cooling}

First, atoms are cooled by isotropic light cooling \cite{PhysRevLett.69.2483,PhysRevA.49.2780,Guillot:01}, the light being injected in the cavity by six single-mode fibres (not in Fig.~\ref{fig_setup}). A $100$~ms long Doppler cooling \cite{Hansch197568,PhysRevLett.40.1639} is performed with about $120$~mW of total power, detuned by $13.6$~MHz ($2.24~\Gamma$) from the $^{87}$Rb cycling transition. Repumping is obtained by phase modulation. This phase is followed by a 2~ms long sub-Doppler cooling stage \cite{Dalibard:89}. In this process the laser beam is detuned by $58.4$~MHz ($9.6~\Gamma$) and the overall power is reduced by $40$~\%. We also reduce the repumper intensity. At the end of the cooling stage we have $\sim 2\times10^{7}$ atoms in all magnetic sub-levels at a temperature around $10~\mu$K.

\subsection{Depumping}

State preparation is achieved by tuning the laser to the $F=2 \rightarrow F'=2$ transition without repumper for 2\,ms. Only the atoms in the $F = 1, m_F = 0$ magnetic sub-level will interact with the micro-wave, the other transitions being Zeeman-shifted by the constant magnetic field of $10$~mG.

\subsection{Interrogation}

%The atomic interrogation is transmitted by a microwave antenna, in the resonant cavity at the $|5S_{1/2},F=1\rangle \rightarrow |5S_{1/2},F’=2\rangle$ clock transition of $^{87}$Rb. The cavity operates a cylindrical TEM$_{011}$ mode with a quality factor of $Q = 6636$. The microwave signal is generated by subtracting the signal of a DDS to the 7 GHz signal from the microwave frequency chain. The DDS and the microwave frequency chain work by frequency multiplication of a 100 MHz reference signal from a commercial quartz oscillator. All these stages are represented on the diagram of the $^{87}$Rb D$_{2}$ transition hyperfine structure on the sub-figure a of the Fig.~\ref{fig_clock}. To get a maximum control of the cavity resonance frequency we control its temperature with a heating coaxial cable.

We work with a Ramsey interferometer scheme~\cite{PhysRev.78.695} in $\pi/2$ -- $\pi/2$ configuration. The microwave which is resonant with the $F = 1, m_F = 0 \rightarrow F = 2, m_F = 0$ clock transition of $^{87}$Rb, is injected into the cavity through one antenna. At this frequency, the cavity sustains a cylindrical TEM$_{011}$ mode with a quality factor $Q = 6636$. To get a maximal stability of its resonance frequency, we control its temperature with a coaxial heating cable winded around it. The microwave pulses are achieved by switching the microwave signal. This gives an atomic phase-shift according to the frequency difference between the atoms clock transition and the local oscillator (see Eq.~\eqref{eq_N}).

\subsection{Detection}

Finally, absorptive detection is performed on resonance with a photodiode and without repumper, with the vertical beam only. The intensity is close to $I_\text{sat}/100$ to prevent any saturation effect (for the $^{87}$Rb D$_{2}$ transition, $I_\text{sat} \simeq 1.67$~mW/cm$^{2}$). A second photodiode measuring the laser's incident intensity is used to normalise the absorption signal and improve the signal-to-noise ratio~\cite{McGuirk:01}. This allows to measuring the number of atoms remaining in $F = 2$ after the micro-wave interaction.

\subsection{Clock operation}

\begin{figure*}
	\centering
	\includegraphics[width=0.8\textwidth]{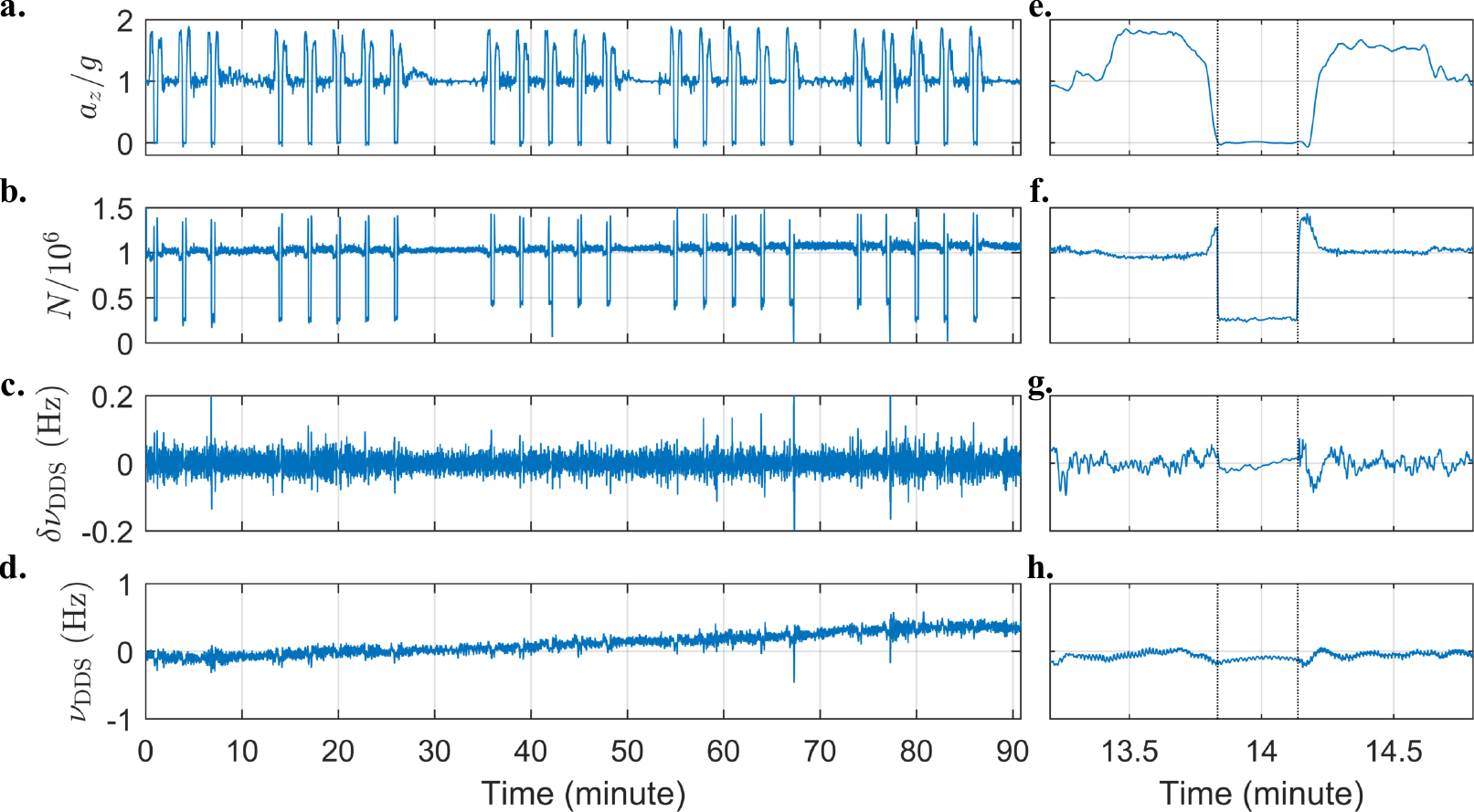}
	\caption{Local oscillator locked to the atomic signal during the entire flight. \textbf{a.} Vertical acceleration in units of $g = 9.81$~m/s$^2$. \textbf{b.} Atom number $N$. \textbf{c.} Frequency correction applied to the local oscillator at 6.835~GHz (through the DDS, see Fig.~\ref{fig_setup} and text). \textbf{d.} DDS frequency (offset of $\sim 165$~MHz removed). \textbf{e, f, g, h.} Zoom on the 4th parabola.}
	\label{fig_clock}
\end{figure*}

In this article, we present results obtained either by scanning experimental parameters (LO frequency, Ramsey time, etc.) or in closed-loop operation. In this latter case, the LO is stabilised to the Rb clock frequency by alternately measuring the atom number at ``half-fringe'' on either side of the central fringe (frequency $\nu \simeq \nu_c \pm \Delta\nu/2$). The difference between two consecutive measurements gives an error signal used to correct the LO frequency. Feedback is applied to the DDS frequency, not directly to the OCXO which remains ``free'' (its frequency may drift). The correction signal and the atom number are shown Fig.~\ref{fig_clock} where the LO is locked to the atomic resonator during an entire flight.

\section{Issues}

Close to resonance, the interference signal is well approximated by:
\begin{equation}
	N(\nu) = N_{1/2}\left[1+C\cos\left(\pi\dfrac{\nu-\nu_{c}}{\Delta\nu}\right)\right]
	\label{eq_N}
\end{equation}
where $N_{1/2}$ is the atom number at half-fringe, $C$ the contrast of the interference pattern, $\nu$ the LO frequency, $\nu_\text{c}$ the $^{87}$Rb clock frequency, and $\Delta\nu \simeq 1/(2 T_{R})$ the full width at half maximum (FWHM) of the central Ramsey fringe. $T_{R}$ is the Ramsey interrogation time. In closed-loop operation, the clock's fractional frequency stability \cite{377790} is related to the signal, noise, contrast, Ramsey time and cycle duration $T_{C}$ by
\begin{equation}
	\sigma_{y}(\tau) = \dfrac{1}{\pi}\dfrac{\Delta\nu}{\nu_{c}}\dfrac{1}{\text{SNR}}\sqrt{\dfrac{T_{C}}{\tau}} ,
	\label{eq_sigma}
\end{equation}
$\tau$ being the integration time. Here, the relevant signal-to-noise ratio is $\text{SNR} = C N_{1/2}/\delta N_{1/2}$, where $\delta N_{1/2}$ is the noise at half fringe, which is well estimated by the first point of the Allan deviation of the atom number at half fringe. For all the data presented, the cycle time is $T_{C} = T_{R} + 130$\,ms (sum of the cooling, preparation and detection times). Therefore, for a given species and transition, the clock is optimized by maximizing the ratio $\text{SNR} \times T_{R}/\sqrt{T_C}$.

On ground, we found this optimum for a cooling and Ramsey times close to $100$~ms and $T_R = 40$~ms respectively (see Fig.~\ref{fig_stability}, red stars). This duration of the cooling stage is a compromise between gathering enough atoms to be shot-noise limited~\cite{PhysRevA.82.033436}, i.e. having a good SNR, and not making the cycle time too long. The optimum of $40$~ms is related to the free fall of the atoms in the cavity (corresponding to $\sim 8$~mm of displacement). When this time becomes longer, the contrast drops dramatically. The temperature is on the order of $10~\mu$K and does not limit significantly on ground.

In weightlessness, one can expect to increase $T_R$ significantly, therefore improving the stability (see Eq.~\eqref{eq_sigma}).
%
%\begin{equation}
%\text{SNR} = C \dfrac{N_{1/2}}{\delta N_{1/2}}
%\label{eq_SNR}
%\end{equation}
%

If we increase the interrogation time $T_{R}$ we reduce the FWHM  of the fringes $\Delta\nu$ and obtain a better frequency discrimination.

%The fractional frequency stability is related to the atom number, contrast, and Ramsey time by
%
%\begin{equation}
%\sigma_{y}(\tau) = \dfrac{1}{\pi}\dfrac{\Delta\nu}{\nu_{c}}\dfrac{1}{\text{SNR}}\sqrt{\dfrac{T_{C}}{\tau}} ,
%\label{eq_sigma}
%\end{equation}
%
%where 
%
%\begin{equation}
%\text{SNR} = C \dfrac{N_{1/2}}{\delta N_{1/2}}
%\label{eq_SNR}
%\end{equation}
%
%is the signal to noise ratio at half fringe, $T_{C}$ the clock's cycle time, and $\tau$ the measurement integration time.

%We therefore define the figure of merit
%%
%\begin{equation}
%m(\tau) = \pi \frac{\text{SNR}}{\Delta\nu} \sqrt{\frac{\tau}{T_{C}}} ,
%\label{eq_figure_of_merit}
%\end{equation}
%%
%which takes into account conflicting effects affecting the stability.

For instance, longer Ramsey times tend to improve the stability because of the narrower fringe, but to worsen it because of the reduction of the atom number and contrast.

%\section{Performances in a laboratory environment}

On Earth we are limited by the falling time of the atoms which leads to dramatic loss of atoms and contrast for an interrogation time longer than 60 ms. So we have made measurements in microgravity.

%\subsection{Performances in-flight and in microgravity}
%
\subsection{Parabolic flights: the environment}

The Rubiclock experiment participated in two parabolic flight campaigns onboard the Airbus A300 ZERO-G. These campaigns were funded by CNES and organized by Novespace. Each campaign consists in three flights and each flight includes 31 parabolas. Thus a whole campaign provides 31 minutes of microgravity, but sliced into 93 slices of 20 seconds. Each microgravity slice is framed by two 20 second slices of 2~g hyper-gravity, all separated by a 1g phase lasting at least 1 minute. Under these conditions, it is not possible to integrate the atomic signal more than 20 seconds at a time during 0~g or 2~g phases.

During the microgravity phases, the residual acceleration felt by the setup depends on its location in the plane, and of course on the parabola quality. The highest fluctuations are in the direction of the local vertical of the aircraft, and are typically about 0.01~g RMS with oscillations of +/- 0.05~g P-V, increasing up to +0.1~g at the beginning and at the end of the parabola. As the detection beam of the setup is only 7~mm diameter, horizontal acceleration fluctuations have a significant effect on the detected atom number. Therefore, the SNR of the atomic signal is all the more limited by these fluctuations as the Ramsey interrogation time is long.

%\begin{figure}
%	\includegraphics{./figs/UTC}
%	\caption{Plot of the atomic-derived correction of the local oscillator during xx minutes of flight.}
%	\label{fig:DDS}
%\end{figure}

%\subsection{On-board operation of the clock}

We manage to operate the clock continuously during the parabolic flight that included the parabolas, wild movements into the Earth's magnetic field and a lot of jolts.

\subsection{Measurements in microgravity}

A trade-off has to be found on the Ramsey time, to optimize the clock frequency stability, which is proportional to 1/($T_{R}$ x SNR). Measurements performed during the first campaign show that the optimized Ramsey time for 0g phases is between $100$ and $200$~ms. Above $200$~ms, the SNR is quickly degraded by the fluctuations of acceleration of the plane. If installed onboard less noisy carriers, as uninhabited satellites, this clock could greatly improve its metrological performance with Ramsey times up to $500$~ms.

During a first campaign we carried out measures of atoms recaptures and measurements of Ramsey fringes with long interrogation time. The phenomenon of recapture is increased in microgravity because the atomic cloud does not fall but thermally expands. We can recapture atoms that remained in the cavity, this allows us, for the same cooling time, to get more atoms.

\begin{figure}[h]
	\includegraphics{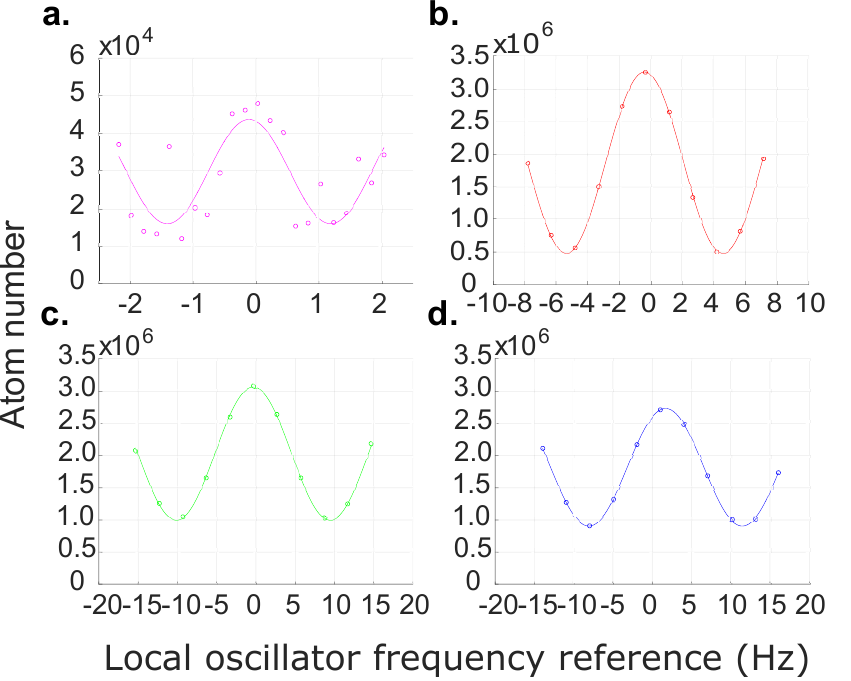}
	\caption{Fringes for different interferometry times and different environment. The subfigure a. represents fringe at $400$~ms in $0$~g. The subfigure b. represents fringe at $100$~ms in $0$~g. The subfigure c. represents fringe at $50$~ms in $1$~g on-board and the subfigure d. represents fringe at $50$~ms in 1g on-ground.}
	\label{fig_fringes}
\end{figure}

Regarding the interrogation time we reached $400$~ms with a contrast of $56$~\%, as we can see on the sub-figure a of the Fig.~\ref{fig_fringes}. This equates to a $1.25$~Hz fringe FWHM. While on Earth we cannot go further than $70$~ms with a contrasts of $29$~\%. This represents a good result for an onboard clock, considering the vibration noise aboard the aircraft.

We participated in a second flight campaign, during which we activated the frequency feedback loop to keep the local oscillator frequency on the rubidium clock frequency. This feedback loop performs an integration over five points. First we tested the clock capacity to stay locked during a flight. During a flight the aircraft performs thirty-one $0$~g phases, each framed by $2$~g phases, all separated by $1$~g phases. We modified the interrogation time between the $0$~g phase and the other phases. The clock stayed locked on the rubidium frequency during all the phases ($31$ $0$~g phases and $62$ $2$~g phases, during more than $90$~mn).

\subsection{Discussion over the results}

%\subsection{Short-term stability estimation}

We now try to estimate the utlimate reachable short-term stability of RubiClock in weightlessness conditions. A direct measurement of the stability showed several issues:

\begin{enumerate}
	\item integration time limited to 20~s,
	\item absence of reference metrological signal for direct comparison,
	\item high sensitivity of the LO to vibration leading to large increase of the Dick effect and degradation of the measured short-term stability,
	\item dependance of the atom number with acceleration fluctuations of the carrier, leading to a short-term stability degradation in noisy environment such as on the aircraft, compared to those which can be reached in quiet conditions like in satellite.
\end{enumerate}

Moreover, to keep the clock frequency locked on the Rb transition although the $1$~g, $2$~g, $0$~g phase successions, the feedback loop had to have a long time constant ($\simeq$1~s), which resulted in a slow frequency drift after each transition from $2$~g to $0$~g corresponding to a relaxation towards a new stationary state.

In this experiment, our LO is not optimised for the vibration, in an aircraft the clock stability is therefore limited by the LO stability because of the vibrations. However, some OCXO have been developed to be less sensitive to the accelerations, like the one for the Pharao clock \cite{Laurent2006}. We get rid of the noise due to the Dick effect \cite{Dick} by performing our noise measurements at the top of the fringe. In this way we are only sensitive to the instrumental and quantum noises of the resonator. This gives us the ultimate stability that we can get with our clock, or whithin a less noisy carrier, like a satelitte.

\begin{figure}[h]
	\includegraphics{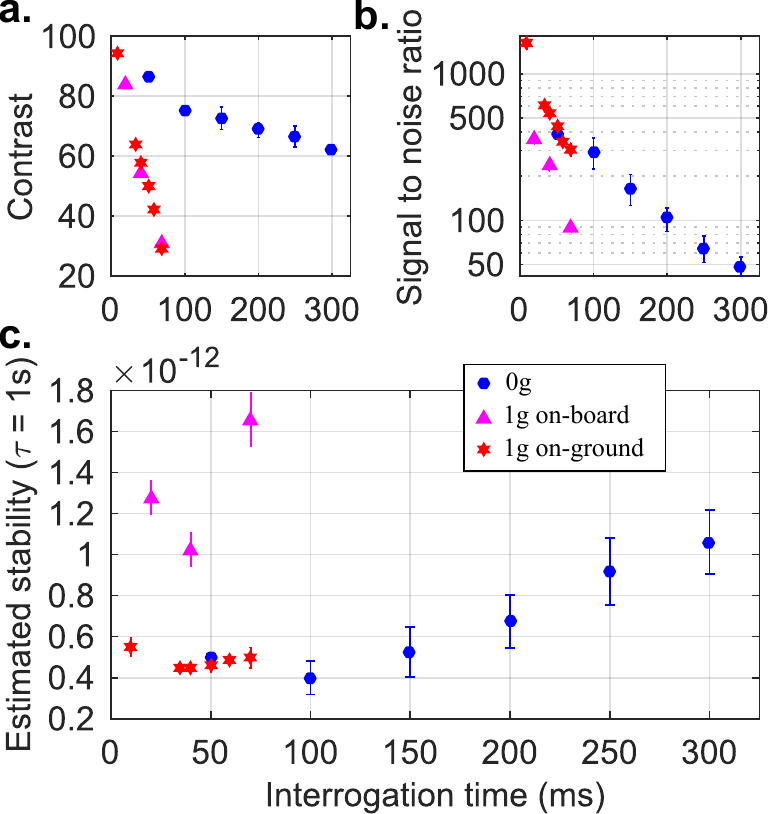}
	\caption{At the top left the contrast of the interferometer. At the top right the signal to noise ratio. At the bottom the estimated stability at 1~s. We perform several measurements for each time, we plot the mean and the standard deviation of this values. The blue circles are the values in 0g. The purple triangles are the values in 1g on-board. The red stars are the values in 1g on-ground.}
	\label{fig_stability}
\end{figure}

\begin{figure}[h]
	\centering
	\includegraphics[width=0.45\textwidth]{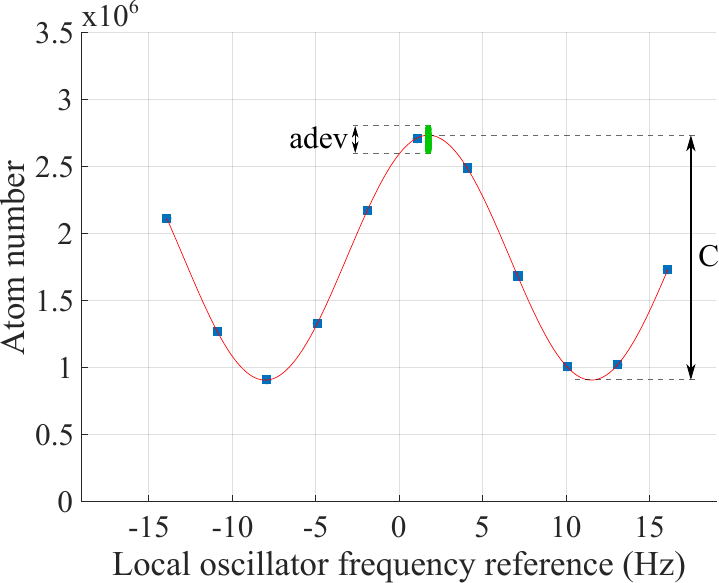}
	\caption{Calculation of the resonator short-term stability. The blue squares are the atom number in function of the interrogation frequency reference. The red line is the sine fit. The green circles are the atom number at the top of the fringe. With the sine fit we compute the fringe contrast and with the atom number at the top of the fringe we compute the Allan deviation.}
	\label{fig_measurement}
\end{figure}

In order to compute the resonator short-term stability $\sigma_{y}(1s)$ we measure the atom number versus the interrogation frequency which we scan with the DDS, as shown Fig.~\ref{fig_measurement}. Since we are limited by the parabola time ($\sim 20$~s) we only take a few points (approximately 10) and perform a sine fit (see Eq.~\eqref{eq_N}). For the noise, we made several measures of the atom number at the top of the fringe (approximately 50). All this is done during the same parabola. This allows to determine all the input parameters (see Eq.~\eqref{eq_sigma1}). The relevant SNR is the inverse of the atom number relative Allan deviation for a delay of one experimental shot (see Eq.~\eqref{eq_adev}). This is because the error signal of the clock feedback loop is computed from two consecutive experimental shots, such that noise occurring at a higher time delay is rejected.

\begin{equation}
	SNR_{1} = \frac{1}{adev}
	\label{eq_adev}
\end{equation}

As we said before:

\begin{equation}
SNR = C SNR_{1}
\label{eq_SNR1}
\end{equation}

Considering the Eq.~\eqref{eq_sigma} we can estimate the stability at 1~s with:

\begin{equation}
	\sigma_{y}(1s) = \frac{1}{\pi}\frac{1}{2T_{R}}\frac{1}{\nu_{0}}\frac{adev}{C}\sqrt{T_{C}}
	\label{eq_sigma1}
\end{equation}

This expression neglects the shot noise and quantum projection noise, which are negligible compared to the noise we measure given our atom number. It also neglects the frequency noise of the LO and the Dick effect which are both expected to be well below the noise we measure by performing the measurement at the top of the fringe.

The estimated stability is plotted in Fig.~\ref{fig_stability} for various experimental conditions (0~g, 1~g on-board and 1~g on-ground) and for $\tau = 1$~s.

We carried out measures in 0~g, and also during the $1$~g phases on the aircraft to see the microgravity improvement. We have performed measures on the aircraft parked on the ground to observe the signal degradation brought by vibration during the flight. We show on Fig.~\ref{fig_stability} the stability measures at 1~s, according to the atoms interrogation time, as well as the contrast and the SNR.

\begin{figure}[h!]
	\includegraphics[width=\linewidth]{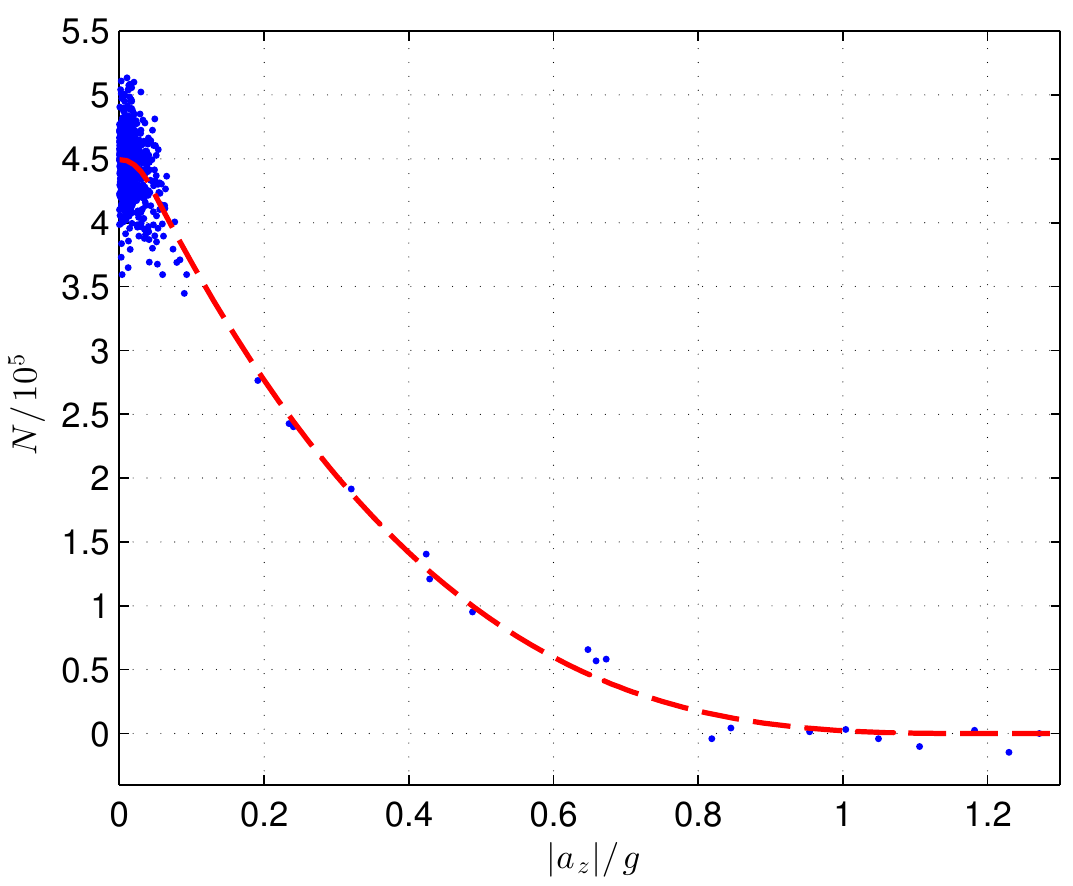}
	\caption{Correlation between vertical acceleration and atom number in closed-loop operation.}
	\label{fig_correlation}
\end{figure}

In $0$~g, the blue circles have a maximum of estimated stability around $100$~ms, with $3.8 \times 10^{-13}$ for the best measurement. But the points are quite dispersed. This is explained by the aircraft vibration, which induces displacement of the atoms during the interrogation time. The standard deviation of the position of the atoms in the quartz bulb is around $0.4$~mm in the vertical axis and $1.6$~mm in the radial axis. This is enough to bring out a few atoms of the $7$~mm detection beam. The atom number depend strongly of the vertival acceleration as we can see on Fig. \ref{fig_correlation}, with variations due to aircraft vibrations. In $1$~g on the aircraft, the purple triangles have a maximum of stability around $40$~ms, with the best measurement at $9.1 \times 10^{-13}$, but on the ground with the same interrogation time the red stars have a stability at 1~s of $4.2 \times 10^{-13}$.

%\subsection{Projected performances in-orbit and in-flight}

\subsection{Projection and comparisons}

In this experiment geometry, with a vertical detection beam, stability strongly depends of the horizontal vibration noise, even if the stability is still sufficient to achieve good performance. As we can see on Fig.~\ref{fig_stability}, the $1$~g on-ground measurements are more than twice as good as the $1$~g on-board measurements, due to the aircraft vibrations, but the $0$~g on-board measurements are still better than the $1$~g on-ground measurements. Considering this, in a low noise $0$~g environment, such as a satellite, we can estimate that this clock will have a short term stability at least twice better than on the aircraft. HORACE, the previous experiment based on the same architecture but with $^{133}$Cs atoms, has a short-term stability of $2.2 \times 10^{-13}$, which is alsmost twice better than our clock on ground, and a long-term stability of $4 \times 10^{-15}$ after $5 \times 10^{3}$~s of integration \cite{PhysRevA.82.033436}. In the future, with more work to reduce the detection noise and identify technical noises during the interrogation (which are currently performed), we can expect to reach less due to the lower cold-collision shift of the $^{87}$Rb \cite{RevModPhys.71.1} and further improve the short-term stability. After consideration of these arguments, in a low noise $0$~g environment and reducing the noise budget like on HORACE, we can hope to reach in a satellite a short-term stability around $10^{-13}$.

Other compact atomic clocks using different approaches are developped. The miniature atomic clock \cite{1574029}, based on MEMS and optoelectronic devices, is more compact but does not reach the same short-term stability, $4 \times 10^{-10}$. The 5071A, a commercial clock based on a cesium beam tube assembly \cite{0026-1394-9-3-002}, can reach a stability of $5 \times 10^{-12}$ in $1$~s in a small volume. Another approach is to use trapped-atom clock on a chip \cite{PhysRevA.92.012106}, which can reach a stability of $5.8 \times 10^{-13}$ in $1$~s. Clock based on coherent population trapping \cite{PhysRevApplied.7.014018} have a short-term stability of $3.2 \times 10^{-13}$ with a small head sensor ($<~1$~L).

All these clocks are good instruments for compact and transportable timebase. Due to its configuration, by isotropic light cooling \cite{PhysRevLett.69.2483,PhysRevA.49.2780,Guillot:01} and cold atoms in free fall in a microwave cavity, Rubiclock experiment is the clock that best benefits from the $0$~g environment as it allows to increase its interrogation time by keeping the same contrast (as shown in Fig.~\ref{fig_stability}.a) and increasing the recapture effect to reduce the cycle time.

\section{Conclusion}

In conclusion we have demonstrated the interesting short-term stability of a compact and transportable atomic clock ($<100$ l), its better performances in microgravity and its ability to operate in a noisy environment. The clock is already in industrial transfer with the company Muquans. Its characteristics make it an interesting candidate for space clocks. Long-term stability measurements have begun to be carried out, to be compared with a comparison with the frequency standard located in SYRTE.

\section*{Acknowledgments}

We are grateful to Bruno Desruelle and the Muquans team, Fabrice Tardif, Mathieu Gu\'{e}ridon, Rapha\"{e}l Bouganne, for their contribution to the laser system, control unit and software, and assistance in designing of the vacuum chamber.
We would like to thank the electronic service of the SYRTE, Michel Lours, Laurent Volodimer and Jos\'{e} Pinto, the mecanical service of the SYRTE, Bertrand Venon, Florence Cornu, Stevens Ravily and Louis Amand, the mecanical service of the GEPI, Jean-Pierre Aoustin, and the mecanical service of the LERMA, Laurent Pelay, for the great support they provide us during the setting up of the clock and the parabolic flight campaign.
We want to thank the CNES, J\'{e}r\^{o}me Delporte, Fran\c{c}ois-Xavier Esnault and Philippe Guillemot for their constant support and their help during the parabolic flight campaign.

%\bibliography{biblio}


\end{document}